\documentclass[preprint,11pt]{elsarticle}
\usepackage[latin1]{inputenc}
\usepackage[english]{babel}
\usepackage{amsmath,amssymb}
\usepackage{epsfig,graphicx}

\psfull

\newcommand\npb[3] {{\it Nucl.\ Phys.\ }{\bf B #1} (#2) #3}
\newcommand\prd[3] {{\it Phys.\ Rev.\ }{\bf D #1} (#2) #3}
\newcommand\prl[3] {{\it Phys.\ Rev.\ Lett.\ }{\bf #1} (#2) #3}

\def\with respect to{with respect to }

\numberwithin{equation}{section}
\begin{document} 
\begin{titlepage}
\begin{flushright}
{MAN/HEP/2011/03}
\end{flushright}     

\begin{center}
{\Large \bf
QCD predictions for new variables to study dilepton transverse momenta at hadron colliders}
\vspace*{1.5cm}

Andrea Banfi$^{a}$, Mrinal Dasgupta$^b$ and Simone Marzani$^b$
\\
\vspace{0.3cm}  {\it
{}$^a$Institute for Theoretical Physics, ETH Zurich, \\
        8093 Zurich, Switzerland\\ \medskip
{}$^b$School of Physics \& Astronomy, University of Manchester,\\
Oxford Road, Manchester, M13 9PL, England, U.K.}\\
\vspace*{1.5cm}

\bigskip
\bigskip

 { \bf Abstract }
\end{center}
\begin{quote}
The D0 collaboration has recently introduced new variables, $a_T$ and $\phi^*$ to more accurately probe the low $Q_T$ domain of $Z$ boson production at hadron colliders than had been previously possible through a direct study of the $Q_T$ distribution. The comparison of such accurate data to precise theoretical predictions from QCD perturbation theory will yield important information on the ability of resummed QCD predictions as well as parton shower models to describe the low $Q_T$ domain and should enable more stringent constraints on non-perturbative effects. In the present Letter we provide analytical predictions for the above mentioned variables, that contain resummation of large logarithms, including next-to--next-to leading logarithmic (NNLL) terms, supplemented by exact $O(\alpha_s^2)$ next-to--leading order calculations from MCFM.
\end{quote}
\end{titlepage}

\section{Introduction}
The transverse momentum distribution of lepton pairs produced via the Drell-Yan process \cite{DY} at hadron colliders, has been one of the most extensively studied variables in QCD~\cite{CS}--\cite{RESBOS}. In spite of a wealth of literature on the subject, measurements of $Q_T$ spectra of lepton pairs (or equivalently of gauge bosons that decay to lepton pairs) and various theoretical approaches to the low $Q_T$ regime still form a significant part of modern QCD phenomenology (for recent theoretical and phenomenological work on the subject see for instance Refs.~\cite{Bozzipt2, petriello, BecNeu}). 

The reason for strong and sustained theoretical interest in the low $Q_T$ 
domain of say $Z$ production at hadron colliders is because this region receives important contributions from the multiple emission of soft and/or collinear gluons. The ability of the theory to make precise predictions for the low $Q_T$ domain, which incorporate the resummation of large logarithms, can thus be taken as a strong signal that the relevant perturbative QCD dynamics is well understood. The low $Q_T$ region is hence also of interest for the purposes of testing, further developing or improving parton shower models embedded in QCD event generators and a comparison between parton showers and analytical resummed predictions is always of interest in this context. 

Moreover, the state of the art for perturbative calculations in QCD is currently such that for variables such as the Drell-Yan and Higgs boson $Q_T$ spectra resummed predictions exist to NNLL accuracy in the resummed exponent, which when combined with NLO estimates from fixed-order calculations ought to produce some of the most precise predictions for QCD observables to date. 

However, while the existence of such accurate resummed calculations is 
encouraging, there are a number of issues that are omitted from the perturbative resummation which could play a significant role in the confrontation of theory with experiment. 
One aspect of physics at hadron colliders, that becomes important at small $Q_T$, is the role of non-perturbative effects commonly attributed to the intrinsic $k_T$ of partons within the proton. One may therefore view any opportunity to compare precise perturbative predictions with accurate experimental data as a chance to place limits on the size of such effects or to effectively parametrise them as for instance in Ref.~\cite{BLNY}. Moreover there are additional issues which may arise such as the role of small-$x$ effects that are neglected by conventional $Q_T$ resummations but may become important at values of $x$ relevant for certain LHC processes such as Higgs production~\cite{BLNYB}. Once again 
the existence or otherwise of such small-$x$ enhanced terms may be inferred from a comparison of conventional resummation with data in the region of large vector boson rapidity (which corresponds to small-$x$ for one of the incoming partons). The need or otherwise, from a phenomenology viewpoint, 
of alternative theoretical constructions such as unintegrated parton densities~\cite{Levin}--\cite{CollinsEllis}
may also be discussed in the context of such a comparison.

Further obscuring the issue of the relevant physics in the low $Q_T$ domain are certain technical issues concerning resummation of logarithms of $M/Q_T$ where $M$ is the hard scale of the process. Most current methods rely on carrying out the resummation in impact parameter ($b$) space and then Fourier transforming to momentum space to obtain the $Q_T$ spectrum. Unfortunately the inversion to $Q_T$ space is not sensible unless one makes certain ad-hoc modifications to the pure resummed predictions. While these modifications are formally subleading from the point of view of the logarithms, they can become numerically important in the $Q_T$ distribution. Clearly such artefacts do not pertain to genuine non-perturbative effects and hence, if significant, potentially obscure a meaningful comparison of theory and experiment. 

In summary the ever increasing accuracy of resummed predictions and the corresponding increased precision of the latest experimental data from hadron colliders as well as the need to use these techniques in the future (for instance in studying the distributions for new particles that may be discovered at the LHC) all contribute to the continued importance of studies of $Q_T$ spectra. 

In our present Letter we address the theoretical prediction, not for the $Q_T$ spectrum itself but consider instead the novel variables introduced in Refs.~\cite{WV,WVBRW} and recently measured by the D0 collaboration~\cite{D0dphi}. These variables labelled the $a_T$ and $\phi^*$ both crucially depend on the azimuthal angle $\Delta \phi$ between the final state leptons, at low $Q_T$. Since it is possible to measure the $a_T$ and $\phi^*$ with significantly lower experimental errors \cite{WVBRW}, in the low $Q_T$ region, an accurate theoretical prediction for these variables, along the line of those for $Q_T$ resummation, is important. The goal of our present Letter is to extend our existing resummed calculation for the $a_T$ distribution~\cite{BanDasDel} in such a manner as to provide a final matched result, taking into account resummation, including  NNLL terms and matching it to fixed order predictions, accounting for the full D0 experimental cuts. 

We organise the present Letter as follows. In the following section we introduce the $\phi^*$ variable measured by the D0 collaboration and relate it in the soft limit to the $a_T$ variable treated in our earlier work~\cite{BanDasDel}. Having done so we provide in the subsequent section the resummed result for both the $\phi^*$ and the $a_T$, incorporating the cuts on the final state leptons as used by the D0 collaboration. We also now include the $\alpha_s^2 L$ (NNLL) term in the resummed exponent and by expanding our results to order $\alpha_s^2$ and comparing with fixed order predictions from the program MCFM~\cite{CamEllis}, demonstrate that we have complete analytical control of all large logarithms that may arise up to the two loop level. This fact allows us to straightforwardly match the fixed-order results from MCFM to the resummed results without the need to provide an ad-hoc prescription for treating subleading logarithmic pieces left over from the fixed order calculations. We 
conclude by pointing out that our matched resummed result can then be used as a basis for detailed phenomenology and comparisons between different theoretical and  Monte Carlo approaches, which is work in progress. 

\section{The $\phi^*$ observable and its relationship to $a_T$}
The $\phi^*$ observable as defined in Ref.~\cite{WVBRW} is 
\begin{equation}
\phi^* = \tan \left (\phi_{\mathrm{acop}}/2 \right) \sin \theta^*, 
\end{equation}
 where $\phi_{\mathrm{acop}} =\pi -\Delta \phi$ and $\Delta \phi$ is the azimuthal angle between the two leptons produced by the $Z$ decay and $\sin \theta^*$ is the scattering angle of the dileptons with respect to the beam in the dilepton rest frame. Two variants of the $\phi^*$ arise by considering two different definitions of $\theta^*$ where one can choose it to either be the angle in the Collins-Soper frame \cite{CS} (which is a particular dilepton rest frame) or one can simply boost along the beam direction such that the leptons make an angle $\theta^*$ and $\pi-\theta^*$ with respect to the beam~\cite{WVBRW}. This definition of $\theta^*$ which avoids the necessity for measuring magnitudes of lepton momenta significantly helps the experimental resolution as explained in Ref.~\cite{WVBRW}. From our viewpoint as far as the resummation is concerned, since we are interested in the low $Q_T$ region, where $Q_T$ is the pair transverse momentum, the various different definitions of $\theta^{*}$ coincide. Let us consider the situation in the dilepton rest frame. In the limit $Q_T \to 0$ the angle $\theta^*$ can be straightforwardly expressed (in terms of lab frame variables or boost invariant quantities) 
as $\sin ^2 \theta^* = 4 l_T^2/M^2$, where $l_T$ refers to the lepton transverse momentum and $M$ is the dilepton invariant mass.

Next we derive the relationship between the $a_T$ variable theoretically studied in Ref.~\cite{BanDasDel} and the $\phi^*$ defined above. 
We consider the variable $\phi_{\mathrm{acop}}/2$ and write 
\begin{equation}
\tan \frac{\phi_{\mathrm{acop}}}{2} = \cot \frac{\Delta \phi}{2} = \sqrt{ \frac{1+\cos \Delta \phi}{1-\cos \Delta \phi}}.
\end{equation} 
Next one notes that 
\begin{equation}
\label{eq:dilep}
Q_T^2 = l_{T1}^2+l_{T2}^2+2 l_{T1}l_{T2} \cos \Delta \phi,
\end{equation}
where $l_{T1}$ and $l_{T2}$ are the magnitude of the lepton transverse momenta. 
In the soft limit, $Q_T \to 0$, we have that $l_{T1}\sim l_{T2}$ and in fact it is easy to see that in this limit the leptons are nearly back-to--back in the transverse plane which results in $|l_{T1}-l_{T2}|= Q_T |\cos \alpha|$ where $\alpha$ is the angle made by the $\vec{Q}_T$ vector with the lepton axis in the transverse plane. Using this in Eq.~\eqref{eq:dilep} we get 
\begin{equation}
Q_T^2 \sin^2 \alpha  \approx 2 l_T^2 \left(1+\cos \Delta \phi \right).
\end{equation}
where we have now set $l_{T1} \approx l_{T2}=l_T$. Also in the soft limit one can approximate $1-\cos \Delta \phi \approx 2$ so that we can eventually write 
\begin{equation}
\tan \frac{\phi_{\mathrm{acop}}}{2} =  \sqrt{ \frac{1+\cos \Delta \phi}{1-\cos \Delta \phi}} \approx \frac{Q_T \sin \alpha}{2 l_T}.
\end{equation}
Recalling the definition of the $a_T$ variable as the component of $\vec{Q}_T$ normal to an axis in the transverse plane which coincides with the lepton axis (in the $Q_T \to 0$ limit) one makes the identification $a_T =Q_T \sin \alpha$ and hence we arrive at 
\begin{equation}
\phi^* = \tan \left (\phi_{\mathrm{acop}}/2 \right) \sin \theta^* \approx \frac{a_T}{M},
\end{equation}
with $M$ the dilepton invariant mass.

Thus the $\phi^*$ variable is straightforwardly related, in the small $Q_T$ 
limit to the $a_T$ variable resummed in our previous work. We can thus use our previous resummation and extend it to include the full experimental cuts for the $\phi^*$ as well as the $a_T$. Having done so we shall then check our resummation against fixed order predictions and confirm full analytical control of large logarithms up to the two-loop level, including the NNLL $\alpha_s^2 \ln \phi^*$ . Finally we provide the matching between the fixed-order corrections calculated up to two-loop level, and our 
resummation which puts our work into a form where phenomenological investigation of the non-perturbative piece can be directly carried out.

\section{Resummed result} 
Here we write down, without deriving it in detail, the resummed result which is a simple extension of the result we derived in our previous paper on the $a_T$~\cite{BanDasDel}. The resummation is based on exponentiation of multiple soft and hard collinear 
radiation along with the factorisation of the observable's phase space. The latter point can be better expressed by noting that in the soft limit $Q_T \to 0$ the $\phi^*$ (we avoid referring to the $a_T$ explicitly from now on) can be expressed in terms of the transverse momenta of multiple soft emissions as $\phi^* \sim |\sum_i \frac{k_{ti}}{M} \sin \phi_i|$ where $\phi_i$ is the angle made by the soft emission $k_i$ with the lepton axis. Thus the condition for several soft gluon emissions to produce values of \mbox{$\tan \left( \phi_{\rm acop}/2 \right) \sin \theta^*$} up to some fixed value $\phi^*$, is simply (writing $k_{ti} \sin \phi_i = k_{yi}$ ) 
\begin{equation}
\Theta \left (\phi^*- \frac{|\sum_i k_{yi}|}{M}\right)= \frac{2}{\pi}\int_0^\infty\frac{db}{b}\sin \left( bM \phi^* \right) \prod_i e^{ibk_{yi}}, 
\end{equation}
which shows the factorisation of the phase space constraint for the $\phi^*$ in terms of the product of contributions from individual gluons. The above constraint on only a single component of the $k_T$, $k_{yi}$, is thus reflected by the presence of the $\sin(bM \phi^*)$ function in contrast to the Bessel function that appears in the $Q_T$ resummation case, which shall account for the absence of a Sudakov peak in the $\phi^*$ distribution, as explained in more detail later.
\begin{equation}
\label{eq:resummed}
\sigma \left( \phi^*,M^2, \cos \theta^*, y \right) = \frac{\pi \alpha^2}{s N_c} 
\int_0^{\infty} \frac{db}{b\pi} \sin \left(bM \phi^* \right) 
e^{-R(\bar{b})} \Sigma \left(x_1,x_2,\cos\theta^*, bM,\mu_f \right)\,,  
\end{equation}
where
\begin{equation}
x_{1,2} = \frac{M}{\sqrt{s}}e^{\pm y} \quad {\rm and} \quad \bar{b}= \frac{b e^{\gamma_E}}{2}\,.
\end{equation}
The resummation is encapsulated by the $b$ integral of the exponential suppression $e^{-R(\bar{b})}$ with the $\sin(bM\phi^*)$ factor having arisen from the one-dimensional momentum conservation constraint as explained above. We have written the result as differential in the dilepton invariant mass $M$ as well as the dilepton rapidity (or equivalently the $Z$ boson rapidity) $y$. In the following the factorisation scale $\mu_f$ will be set equal to $M$.

The function $\Sigma$ has an identical structure to the Born level result, where $\theta^*$ is the scattering angle in the dilepton rest frame. Note that $\Sigma$ acquires a dependence on the impact parameter $b$ because of the resummation of logarithms of $b$ via DGLAP evolution, which then determines the scale of the parton distribution functions embedded in $\Sigma$. Including the contributions from the $Z$ as well as from the virtual photon, we  have 
\begin{eqnarray} \label{sigma}
\Sigma &=& (1+\cos^2 \theta^*)\left(Q_q^2- 2 Q_q V_l V_q \chi_1 +(A_l^2+V_l^2)(A_q^2+V_q^2) \chi_2\right)  {\mathcal{F}}_q^+
 \nonumber \\  &&+  \cos \theta^* (-4 Q_q A_l A_q \chi_1+ 8 A_l V_l A_q V_q \chi_2)  {\mathcal{F}}_q^{-},
\end{eqnarray}
where a sum over quark flavours $q$ is implied. The above equation is naturally written as the sum of the two terms with different angular dependence: the first one is proportional to $(1+\cos^2 \theta^*)$ and represents the parity conserving piece of the electro-weak interaction, while the term involving $\cos \theta^*$ is the parity violating piece. We notice that upon integration over the full $\theta^*$ range, as well as over symmetric intervals, the parity violating term vanishes.
The coefficients $A_{l,q}$ and $V_{l,q}$ are the electroweak couplings for lepton $l$ and parton $q$, explicitly given by:
\begin{equation}
A_f = T^3_f \quad {\rm  and } \quad V_f = T^3_f- 2 Q_f \sin^2 \theta_W, \quad f= l,q \,,
\end{equation}
where $T^3_f$ is the third component of the isospin.
 We also have introduced
\begin{eqnarray}
\chi_1 &=& \kappa \frac{M^2 \left(M^2-M_Z^2 \right)}{(M^2-M_Z^2)^2 +\Gamma_Z^2 M_Z^2}, \nonumber \\
\chi_2 &=& \kappa^2 \frac{M^4}{(M^2-M_Z^2)^2 +\Gamma_Z^2 M_Z^2},  \nonumber \\
\kappa  &=& \frac{\sqrt{2} G_F M_Z^2}{4 \pi \alpha}.
\end{eqnarray}

In Eq.~(\ref{sigma}), $\mathcal{F}^{\pm}$ are explicitly given by
\begin{equation}
\label{eq:Fdef}
{\mathcal{F}}_q^{\pm}= \left(\bf{C} \otimes \bf{f}_1\right)_q(x_1,\bar{b})\left(\bf{C} \otimes \bf{f}_2\right)_{\bar{q}}(x_2,\bar{b}) \pm \left(\bf{C} \otimes \bf{f}_1\right)_{\bar{q}}(x_1,\bar{b})\left(\bf{C} \otimes \bf{f}_2\right)_{q}(x_2,\bar{b}).
\end{equation}

The convolutions involving the matrix of coefficient functions $\bf{C}$ and 
the vector of parton densities $\bf{f}_{1,2}$ for incoming hadrons $1$ and $2$ respectively can be explicitly written as 
\begin{equation}
\label{eq:conv}
\left(\bf{C} \otimes \bf{f}_i\right)_q(x_i,\bar{b})= \int_{x_i}^{1} \frac{dz}{z} C_{q \alpha} \left (\alpha_s \left( \frac{1}{\bar{b}} \right),\frac{x_i}{z}\right)
f_i^\alpha\left(z,\frac{1}{\bar{b}}\right),
\end{equation}
where $i=1,2$ and a sum over all flavours $\alpha$ is implied. 

The coefficient functions $C$ represent perturbative corrections to the 
collinear branching of an incoming parton $\alpha$ to a parton $q$ which annihilates with $\bar{q}$ to form the $Z$ boson. We note that the collinear enhanced terms generated by such a branching are incorporated to our accuracy into the scale of the pdfs $f_i$ via their dependence on the impact parameter $b$. Thus the coefficient functions represent only the non-logarithmic constant terms. For our purposes we shall only need the coefficient functions up to first order in $\alpha_s$:
\begin{equation}
C_{q \alpha}(x,\bar{b}) = \delta_{q \alpha}\delta \left(1-x \right)+\frac{\alpha_s(1/\bar{b})}{2 \pi}C^{(1)}_{q \alpha}\left (x \right ) +\mathcal{O} \left(\alpha_s^2 \right).
\end{equation}
The explicit expression for the coefficient functions is given in Appendix~\ref{app:integrals}.

We note that the argument of $\alpha_s$ to be used while evaluating $C_1$ should be of the order of the hard scale $M$ if one works to NLL accuracy but is ${1}/{\bar{b}}$ if one wishes to achieve the NNLL accuracy we seek, hence we have retained an explicit $b$ dependence of the coefficient functions. Other than the $b$ dependence indicated in the coefficient functions and the pdfs the large logarithms in $b$ are resummed into the exponential function involving $R(b)$. 

The resummation of logarithms in $b$ space turns out to be precisely as for the case of $Q_T$ resummation~\cite{BanDasDel} and hence the function $R(\bar{b})$ is the same as for that case. $R(\bar{b})$ can be expressed in the general form \cite{CTTW}
\begin{equation}
R(\bar{b}) = L g_1 (\alpha_s L) +g_2 (\alpha_s L) + \alpha_s g_3 (\alpha_s L) +\cdots
\end{equation}
with $L = \ln \left(\bar{b}^2 M^2 \right)$. The functions $g_1$ and $g_2$ describe leading and next-to--leading logarithms respectively while $g_3$ refers to NNLL terms. A calculation by the Florence group \cite{FlorenceQT, Bozzipt2} extends up to the $g_3$ term but a recent study by Becher and Neubert indicates that one of the assumptions involved in computing part of the $g_3$ piece used in $Q_T$ resummation, namely that the three loop coefficient $A^{(3)}$ is identical to the one appearing in threshold resummation, may in fact not be valid \cite{BecNeu}.  Since the piece of $g_3$ we refer to above starts at order $\alpha_s^3$, for the present Letter we shall avoid this issue pending further clarification by retaining only the well established $\alpha_s^2 L$ piece of $g_3$ which has been known for several years \cite{DS}.  We anticipate that the terms we neglect here corresponding to higher order contributions to $g_3$ will in any case have a negligible effect, but note that it is trivial to include them if required\footnote{A similar conclusion was reached in a revised version of Ref.~\cite{Bozzipt2}, where it was found that the numerical impact of the modified $A^{(3)}$ is of the order of a few percent at small $Q_T$.}.
Explicit expressions for the functions appearing in the radiator are collected in~\ref{app:integrals}.

The D0 collaboration recently measured the $\phi^*$ distribution with a set of cuts on the lepton momenta. More specifically, in the case of muons,  they require the rapidity of both leptons to be $|\eta_{1,2}| < 2$ and transverse momenta $p_T$ of both leptons to be above 15 GeV. These cuts which leave the basic form of the resummation unchanged can be easily incorporated into our calculations by working out the appropriate limits for the $\theta^*$ and $y$ integrals and we have included them in our final result.

In the following section we shall check the resummation by considering its expansion to order $\alpha_s^2$ and comparing with exact fixed-order 
results from MCFM \cite{CamEllis}. We shall demonstrate that the resummation allows us analytical control over all logarithmic terms at $\alpha_s^2$ 
accuracy. 

\section{Comparison to fixed-order calculations}
To expand the resummation to order $\alpha_s^2$ it is most convenient to consider the Mellin moments with respect to $x_1$ and $x_2$. Then one can write the pdfs at scale $1/\bar{b}$ in terms of the pdfs at scale $M$ using DGLAP evolution, which amounts to replacing the convolutions in Eqs.~\eqref{eq:Fdef} and \eqref{eq:conv} by $N$-space products as follows
\begin{equation}
\label{eq:mellin}
\left(\bf{C} \otimes \bf{f}_i\right)_q(x_i,\bar{b}) \to  \left \{ \tilde {\bf{C}}\left (\alpha_s \left( \frac{1}{\bar{b}} \right), N_i\right) \exp \left [ -{\bf{\Gamma}}(\alpha_s,N_i) L\right] \tilde{\bf{f}} \left(N_i,M \right) \right \}_q,
\end{equation}
where $\tilde {\bf{C}}$ are the Mellin moments of the coefficient functions ${\bf{C}}$, while $\alpha_s$ in the argument of the two-loop anomalous dimension ${\bf{\Gamma}}$ is evaluated at scale $M$.  
We also need to expand $\alpha_s \left(1/\bar{b} \right)$:
\begin{equation}
\alpha_s \left( \frac{1}{\bar{b}} \right) = \alpha_s(M)+\beta_0 \alpha_s^2(M) L +O\left(\alpha_s^2 \right).
\end{equation}
The radiator $R(\bar{b})$ can also be expanded up to the relevant $\alpha_s^2$ in the form
\begin{equation}
R(\bar{b}) = \frac{\alpha_s}{2\pi} \left (G_{12} L^2 +G_{11} L \right)+ \left(\frac{\alpha_s}{2\pi} \right)^2\left(G_{23} L^3+G_{22} L^2+G_{21} L \right)+O\left(\alpha_s^2 \right),
\end{equation}
where the coefficients $G_{nm}$ are reported in~\ref{app:integrals}. With the above information in place we can expand the result out to order $\alpha_s^2$ straightforwardly and carry out the $b$ integrals term by term to express the answer in $\phi^*$ space. We also invert (in this case possible by inspection) the Mellin transform from $N_i$ space back to $x_i$ space. 

While the expressions for the expansion of the resummation up to order
$\alpha_s^2$ is too cumbersome to report here, we shall adopt the
strategy of our previous work where we write down an expression for
the difference between the integrated $\phi^*$ and the very well known $Q_T$ distributions.
Because the logarithms in $b$ space are the same in the two cases and because the sine and the Bessel function transforms have little effect on the expansion of the resummation, the logarithms in $\phi^*$ mostly cancel the ones in $Q_T$ and the difference of the distributions
admits a relatively simple analytical form:
\begin{multline} \label{diff_NNLL}
  \tilde{\sigma}\left(N_1,N_2,\phi^*\right)\big |_{\phi^*=\epsilon}-\tilde{\sigma} \left(N_1,N_2,Q_T/2\right) \big |_{Q_T/2=\epsilon}=  \tilde{\sigma}_0 \left(N_1,N_2 \right) \\
  \left(\frac{\alpha_s}{2\pi} \right)^2\left[\pi^2 C_F^2 \ln^2 \frac{1}{\epsilon^2} +\left(-24 C_F^2 \zeta(3)-3 \pi^2 C_F^2 - \frac{4}{3} \pi^3 C_F \beta_0 \right. \right. \\
  \left.\left.+\pi^2 C_F \frac{\left[{\bf\Gamma_0}(N_1)
          {\bf\tilde{f}_1}(N_1)\right]_q {\bf f_2}_{\bar{q}}(N_2)+ 1
        \leftrightarrow 2}{{\bf f_1}_q(N_1){\bf f_2}_{\bar{q}}(N_2)+1
        \leftrightarrow 2}\right) \ln \frac{1}{\epsilon} \right],
\end{multline}
where only the first term involving $\ln^2 \left( \frac{1}{\epsilon^2}
\right)$ had been checked against MCFM in Ref.~\cite{BanDasDel} for the case of the $a_T$ distribution. In
the above equation $\tilde{\sigma}(N_1,N_2)$ represents the Mellin
transformed cross-section, taking Mellin moments with respect to $x_1$ and $x_2$
while $\sigma_0$ is the Born cross-section. $\bf\Gamma_0$ denotes the
leading order anomalous dimension matrix and we have used the matrix
notation we introduced earlier in Eq.~\eqref{eq:Fdef}.

\begin{figure}
  \begin{center}
    \epsfig{file=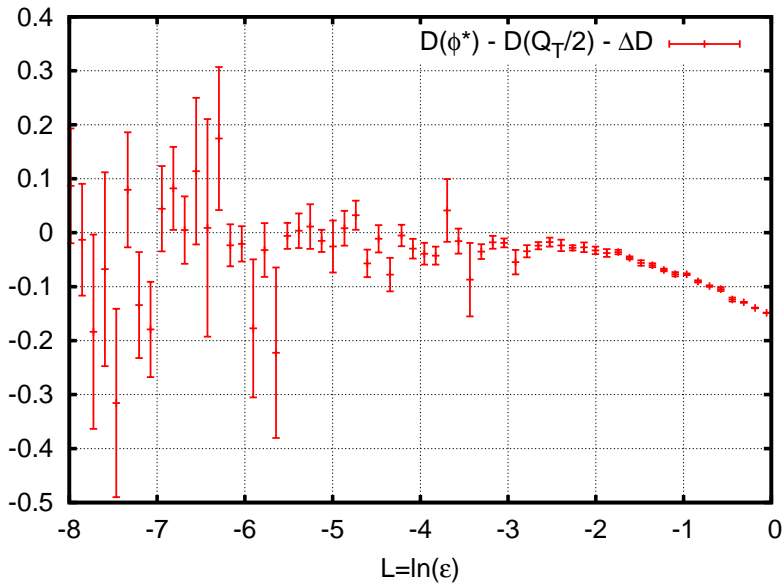,width=0.6 \textwidth }
  \end{center}
\caption{The difference between the NLO differential distributions for $\phi^*$ and $Q_T/2$ from MCFM after removal of logarithmic terms from the resummation.} \label{fig:checklogs}
\end{figure}

By taking the derivative of Eq.~(\ref{diff_NNLL}),  we can easily compute the difference between the logarithmic terms in the differential distributions:
\begin{equation}
\Delta D(\epsilon) = \frac{1}{\sigma_0} \frac{{\rm d }}{{\rm d} \ln \epsilon} \left( {\sigma}\left(\phi^*\right)\big |_{\phi^*=\epsilon}-{\sigma} \left(Q_T/2\right) \big |_{Q_T/2=\epsilon} \right).
\end{equation}
Subtracting $\Delta D$ from the corresponding fixed-order differential distribution $D\left(\phi^*\right)-D\left(Q_T/2\right)$, computed with MCFM at NLO, we should find a result that tends to zero at small $\epsilon$. This difference is plotted in
Fig.~\ref{fig:checklogs} and one notes that the result tends to zero as
expected.
Although we do not report them here, we have explicitly checked the expansion of the resummation against MCFM for the $\phi^*$, the $a_T$  and  the $Q_T$ distributions separately.
\section{Matching to fixed-order}
Having demonstrated complete control over the divergent pieces at
order $\alpha_s^2$ we are in a position to propose a particularly
simple matching formula for the differential $\phi^*$ distribution, where one simply adds the resummation to MCFM
results and subtracts the expansion of the resummation to order
$\alpha_s^2$. This scheme is known as $R$-matching \cite{CTTW} and one
does not need here to provide an arbitrary prescription to treat left
over divergent pieces from the difference between MCFM and the
expanded resummation.

Before we can match the resummed result to fixed-order we have however
to deal with the evaluation of the $b$ integral in
Eq.~\eqref{eq:resummed}. As is well known, the radiator $R(b)$, which
resums terms logarithmic in $b$, is divergent both at small $b$ and
large $b$. The large $b$ divergence is due to the Landau pole in the
running coupling and is associated to non-perturbative behaviour. We
shall hence impose a cut-off at a value $b_{\mathrm{\max}}$, chosen in
the vicinity of the Landau pole, to avoid this problem.
Additionally we need to handle the divergence at small $b$. Since the small $b$ region describes large $Q_T$ or $\phi^*$ it is beyond the 
jurisdiction of our resummation formula.  We decide to freeze the radiator $R(\bar{b})$ for values of $\bar{b} M< 1$, which is the region of large $\phi^*$ not controlled by the resummation~\cite{eec}. 
Different prescriptions, such as the one adopted by the Florence group, where  one modifies the logarithms of $b$ in $R(b)$, $\ln\left(\bar{b}^2M^2 \right) \to \ln\left(1+\bar{b}^2M^2 \right)$ will be investigated in a future phenomenological study. 
Our matched result corresponds to
\begin{equation}
\left(\frac{ {\rm d} \sigma}{{\rm d} \phi^*}\right)_{\mathrm{matched}} = \left(\frac{{\rm d} \sigma}{ {\rm d} \phi^*}\right)_{\mathrm{resummed }} +\left(\frac{{\rm d} \sigma}{{\rm d} \phi^*}\right)_{\mathrm{MCFM}}-\left(\frac{{\rm d} \sigma}{{\rm d} \phi^*}\right)_{\mathrm{expanded}}
\end{equation}
where we have indicated by $ \left( \frac{{ \rm d} \sigma}{{\rm d} \phi^*}\right)_{\mathrm{expanded}}$ the expansion of the resummation to NLO which is achieved by expanding the $b$ space radiator to that accuracy and performing the $b$ integral precisely as for the resummation. For the matching to be considered successful we ought to observe that the matched curve follows the resummed result at small $\phi^*$ while at large $\phi^*$ one expects the resummation to largely cancel  against its expansion (up to relatively small terms varying as $\alpha_s^3$) and hence the result should follow the NLO MCFM curve.

\begin{figure}
  \begin{center}
    \epsfig{file=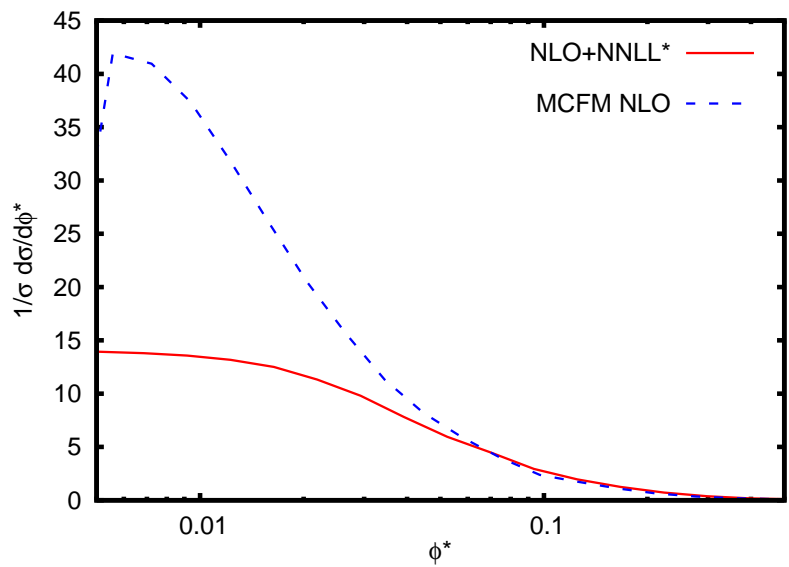,width= 0.6\textwidth}
  \end{center}
\caption{The differential $\phi^*$ distribution computed at NLO with MCFM (dashed blue line) and our final result (solid red) obtained by matching our resummation, which contains some of the next-to next-to leading logarithms and hence is denoted by ${\rm NNLL}^*$, to the fixed order calculation at $O(\alpha_s^2)$, from MCFM. The distributions are normalised to NLO cross section.}  \label{fig:matching}
\end{figure}

Our final result for the matched differential  $\phi^*$ distribution is plotted in Fig.~\ref{fig:matching}, together with the pure fixed order $O\left(\alpha_s^2 \right)$ calculation obtained from MCFM. We notice that the NLO calculation diverges in the region of small $\phi^*$, while the resummed and matched results tend to a constant. This behaviour can be explained by performing a simple calculation at the double logarithmic accuracy:
\begin{eqnarray}
\frac{1}{\sigma} \frac{{\rm d} \sigma}{{\rm d }  \phi^*} &\simeq& \int_0^{\infty} d (b M) \cos(b M \phi^*) e^{- \frac{\alpha_s C_F}{2 \pi} \ln^2 \left({b}^2 M^2 \right)}=
\int_0^{\infty} d (b M)e^{- \frac{\alpha_s C_F}{2 \pi} \ln^2 \left({b}^2 M^2 \right)}\left[1+ O\left({\phi^*}^2\right) \right] \nonumber \\ &=&
\frac{\pi}{\sqrt{2 \alpha_s C_F}}  e^{\frac{\pi}{8 \alpha_s C_F}} + O\left({\phi^*}^2\right).
\end{eqnarray}
The above calculation indicates that the differential distribution goes to a constant at small $\phi^*$. In order to more accurately obtain the value of the asymptotic constant of the $\phi^*$ distribution,  one needs to include the full calculation up to at least NLL terms in the exponent, as we have done here,  as well as include non-perturbative effects. 

While low values of $\phi^*$ or $Q_T$ can be obtained via Sudakov suppression or kinematical cancellation,  in the present case of $\phi^*$ the kinematical cancellation starts to dominate prior to the formation of the Sudakov peak, in contrast to the $Q_T$ case.
This heralds the breakdown of the logarithmic hierarchy in the resummed exponent in the region corresponding to large values of $b$. Hence, a precise evaluation of the $b$ integral becomes important, which however depends on a number of prescriptions referred to earlier. It would be interesting to study the implication of this observation for the phenomenology of non-perturbative effects and to compare the conclusions to the $Q_T$ case, as we intend to do in forthcoming work.

From Fig.~\ref{fig:matching} we note that the matched curve tends to MCFM at large $\phi^*$ but deviates from fixed-order significantly at lower $\phi^*$ and follows the resummed curve as anticipated. We also note that the deviation from NLO will be significant over a large range in $\phi^*$ where there are accurate data and hence we would see a visible role for both resummation and potentially non-perturbative effects in phenomenological studies. 

\section{Conclusions and Outlook}
In this Letter we have presented a resummed result for the newly introduced $\phi^*$ variable, closely related to the previously studied $a_T$ variable, which is accurately measured  down to very low values of $\phi^*$ by the D0 collaboration. Our resummation is valid to full NLL accuracy but we also include the leading order $\alpha_s^2 L$ term of the NNLL resummation which simplifies the matching procedure to a large extent. We note that it is straightforward to include the full NNLL resummation and non-perturbative effects via a Gaussian smearing in $b$ space which developments we leave to our forthcoming phenomenological study.  

For phenomenological purposes a number of further studies of the resummed distribution still require to be carried out. Primarily one needs to ascertain the stability of the resummation against the change of various parameters and prescriptions involved in generating the resummed and matched result. Amongst these parameters are just the traditional factorisation and renormalisation scales which here have been taken to be all equal and set to the invariant mass of the lepton pair. Additionally here one can explore the role of changing the resummation scale or in other words rescaling the resummation variable $\bar{b}M$ to assess the impact of subleading logarithms omitted by the resummation. However, with inclusion of full NNLL effects we may anticipate subleading logarithms to be essentially invisible. 

More importantly there are certain ad-hoc prescriptions inherent in $b$ space resummation such as the treatment of the radiator at large and small $b$ to avoid respectively the Landau pole and the spurious small $b$ divergences beyond the control of resummation. For a complete phenomenological study therefore we intend to produce an uncertainty band on the resummed result which reflects the variation with respect to the change in these parameters and prescriptions. Once this is done we can focus on pinning down non-perturbative effects most accurately, for we which we shall also carefully explore the different rapidity ($y$) bins over which data exist. 

We believe that our accurate theoretical calculations will encourage the use of these novel variables, namely $\phi^*$ and $a_T$, for experimental studies of the $Q_T$ spectrum of the $Z$ boson at the LHC.
From a theoretical viewpoint it is worth noting that the study here is closely related to variables such as $\Delta \phi$ between jets measured recently by ATLAS~\cite{ATLAS_ang} and CMS~\cite{CMS_ang} and computed in Ref.~\cite{BanDasDel08}; those calculations therefore signify an extension of $Q_T$ resummation to processes with colour (jets) in the final state. 
Also for the future it will be interesting to compare and contrast different theoretical approaches including both the newly developed SCET approach \cite{petriello} as well as Monte Carlo parton shower based predictions in the context of the $\phi^*$ variable and the D0 data.
\vspace{1 cm}
 
{\bf Acknowledgments}
We thank Terry Wyatt and Mika Vesterinen for many useful discussions. One of us (MD) gratefully acknowledges the IPPP (Durham) for his associateship award which provided funding for the visit of AB to Manchester where this work was finalised. AB acknowledges the hospitality of the Manchester particle physics group during the completion of this work.
The work of SM is supported by UK's STFC. This research is also supported by the Swiss National Science Foundation under
contract SNF 200020-126632.
\newpage

\appendix

\section{Explicit resummation formulae} \label{app:integrals}
The radiator used in the resummation admits the following expansion:
\begin{equation}
 R(b) = L g_1 (\alpha_s L ) + g_2 (\alpha_s L)+ \alpha_s g_3(\alpha_s L),
\end{equation}
\begin{align}
  g_1(\lambda) =& \frac{C_F}{\pi\beta_0\lambda} \left
    [-\lambda-\ln{(1-\lambda)}\right]\,,\\
  g_{2}({\lambda})=&  \frac{3C_F}{2\pi \beta_0} \ln(1-\lambda)
  +\frac{KC_F [\lambda + (1-\lambda)\ln(1-\lambda)]}
  {2\pi^2\beta_0^2(1-\lambda)} -\frac{C_F \beta_1}{\pi\beta_0^3}
  \left[ \frac{\lambda + \ln (1-\lambda)}{1-\lambda} + \frac{1}{2}
    \ln^2{(1-\lambda)} \right]\,,\\
g_3(\lambda) =& \frac{C_F^2}{\pi \beta_0}\frac{4\pi^2-48 \zeta(3)-3}{16}\frac{\lambda}{1-\lambda} + \dots
\label{eq:g2}
\end{align}
with $\lambda =  \alpha_s(M^2)\,\beta_0 L$, $L= \ln \bar{b}^2M^2$ and $K=C_A \left(\frac{67}{18}-\frac{\pi^2}{6} \right)-\frac{5}{9} n_f$. We recall the reader that in this Letter we  only keep the term in the NNLL coefficient $g_3$ which gives rise to single logarithms at $\alpha_s^2$.
At the aimed accuracy we need to consider the running of the strong coupling constant at two-loops:
\begin{equation}
  \label{eq:twoloop-as}
  \alpha_s(k_t^2) = 
  \frac{\alpha_s(M^2)}{1-\rho}\left[1-\alpha_s(M^2)\frac{\beta_1}{\beta_0}
    \frac{\ln(1-\rho)}{1-\rho}\right] \,,\qquad
  \rho = \alpha_s(M^2) \beta_0 \ln\frac{M^2}{k_t^2}\,,
\end{equation}
where the coefficients of the QCD $\beta$-function are defined as
\begin{equation}
\beta_0 = \frac{11 C_A-2 n_f}{12 \pi}, \, \, \qquad \beta_1 =\frac{17 C_A^2-5C_A n_f -3 C_F n_f}{24 \pi^2}.
\end{equation}
 The function $R$ admits the following perturbative expansion:
\begin{equation}
R(\bar{b}) = \frac{\alpha_s}{2\pi} \left (G_{12} L^2 +G_{11} L \right)+ \left(\frac{\alpha_s}{2\pi} \right)^2\left(G_{23} L^3+G_{22} L^2+G_{21} L \right)\,,
\end{equation}
with the coefficients given by
\begin{align}
 G_{12}= & \,C_F  \\ \,
G_{11}= & -3 C_F\\ \,
G_{23}= & \,C_F\frac{4}{3} \pi \beta_0\\ \,
G_{22}= & \,C_F( K-3 \pi  \beta_0)\\ \,
G_{21}= & \,
C_F^2 \left( \pi^2-12  \zeta(3)  - \frac{3 }{4} \right)+
C_F C_A \left(6\zeta(3) - \frac{193}{12} + \frac{11}{9} \pi^2 \right)  \nonumber \\
&  + C_F n_f \left( \frac{17 }{6}  - 
  \frac{2}{9} \pi^2  \right)\,.
\end{align}
The explicit form of the coefficient functions is~\cite{CSS}
\begin{align}
 C_{q\bar{q}}(\alpha_s(1/\bar{b}),x) = & \delta(1-x) + \frac{\alpha_s(1/\bar{b})}{2 \pi} C_F\left( \left(\frac{\pi^2}{2}-4\right) \delta(1-x)+ (1-x)\right)+O\left(\alpha_s^2 \right)\,, \\
C_{qg}(\alpha_s(1/\bar{b}),x) = &  \frac{\alpha_s(1/\bar{b})}{2 \pi} x (1-x) + O\left(\alpha_s^2 \right)\,. \\
\end{align}

The matching to the fixed-order result computed with MCFM is most easily performed at the level of the differential distribution. Hence, the $b$-integral we wish to perform are of the following form: 
\begin{equation}
\frac{\partial}{ \partial \ln \phi^*} \frac{2}{\pi} \int_0^{\infty} \frac{d b}{b} \sin( b M \phi^*)\ln^m \left( \bar{b}M\right)^2 =
 \phi^* \frac{2}{\pi} \int_0^{\infty} d (b M) \cos( b M \phi^*)\ln^m \left( \bar{b}M\right)^2 \,.
\end{equation}
Explicit results for $m=1,2,3,4$ and $\phi^*>0$ are given by
\begin{eqnarray}
\frac{2}{\pi} \int_0^{\infty} d (bM ) \phi^* \cos( b M \phi^*) \ln \left( \bar{b} M\right)^2 &=& -2, \nonumber \\
\frac{2}{\pi} \int_0^{\infty} d (bM) \phi^* \cos(b M\phi^*) \ln^2 \left(\bar{b} M\right)^2 &=& 4 \ln \left(4 {\phi^*}^2 \right), \nonumber \\
\frac{2}{\pi} \int_0^{\infty} d ( bM) \phi^* \cos( b M\phi^*) \ln^3 \left(\bar{b} M\right)^2 &=& -6 \ln^2 \left(4 {\phi^*}^2 \right)-2 \pi^2, \nonumber \\
\frac{2}{\pi} \int_0^{\infty} d ( bM ) \phi^* \cos( bM \phi^*) \ln^4 \left(\bar{b} M\right)^2 &=& 
8 \ln^3 \left(4 {\phi^*}^2 \right)+8 \pi^2 \ln \left(4 {\phi^*}^2 \right)+128 \zeta(3).  \nonumber \\
\end{eqnarray}

\end{document}